\documentclass[conference]{IEEEtran}
\usepackage{cite}
\usepackage{amsmath,amssymb,amsfonts}
\usepackage{graphicx}
\usepackage{subfigure}
\usepackage{textcomp, color}
\usepackage{tikz}
\usetikzlibrary{decorations.pathreplacing}
\usetikzlibrary{shapes}
\usepackage[algoruled, linesnumbered]{algorithm2e}

\def\BibTeX{{\rm B\kern-.05em{\sc i\kern-.025em b}\kern-.08em
    T\kern-.1667em\lower.7ex\hbox{E}\kern-.125emX}}
\begin{document}

\title{Deep Q-Learning for Dynamic Reliability Aware NFV-Based Service Provisioning}

\author{Hamed Rahmani Khezri $^\dagger$, Puria Azadi Moghadam$^\dagger$, Mohammad Karimzadeh Farshbafan$^\dagger$, \\ Vahid Shah-Mansouri$^\dagger$,
Hamed Kebriaei$^\dagger$, and Dusit Niyato$^+$ \\
$^\dagger$ School of Electrical and Computer Engineering, Faculty of Engineering, University of Tehran \\
$^+$ School of Computer Science and Engineering, Nanyang Technological University \\
emails: \{hamed.rahmani\_7, puria.azadi.m,  m.karimzadeh68, vmansouri, kebriaei\}@ut.ac.ir, dniyato@ntu.edu.sg\}
}
\maketitle

\begin{abstract} Network function virtualization (NFV) is referred to the technology in which softwarized network functions virtually run on commodity servers. Such functions are called virtual network functions (VNFs). A specific service is composed of a set of VNFs. This is a paradigm shift for service provisioning in telecom networks which introduces new design and implementation challenges. One of these challenges is to meet the reliability requirement of the requested services considering the reliability of the commodity servers. VNF placement which is the problem of assigning commodity servers to the VNFs becomes crucial under such circumstances. To address such an issue, in this paper, we employ Deep Q-Network (DQN) to model NFV placement problem considering the reliability requirement of the services. The output of the introduced model determines what placement will be optimal in each state. Numerical evaluations show that the introduced model can significantly improve the performance of the network operator.
\end{abstract}

\section{Introduction}
\label{Section: Introduction}
Service deployment in traditional enterprise networks tightly depends on specific hardware named middlebox [1], [2]. Quality of service (QoS) monitoring tools, video transcoders, firewalls, intrusion detection systems, proxies, and deep packet inspection are examples of such middleboxes. This function implementation limits the expansion of the networks and increases CAPEX and OPEX [3]. Because of these shortcomings in using middleboxes, a fundamental change for network function implementation is inevitable. Network Function Virtualization (NFV) promises to obviate the limitation of middleboxes for deploying new network services [4], [5]. In the NFV framework, the hardware middleboxes are replaced by the modules of software named virtual network functions (VNFs) running on commodity servers [6]. To provide a network service, a set of appropriate VNFs should be sequenced in a chain called service function chain (SFC). The VNFs of a service can be deployed by launching a VM instance in any server of network infrastructure. The procedure of assigning servers to the VNFs of service is named NFV placement.

There are three main components for the NFV based network. The first one is the \emph{services} which are requested by the network users. Each incoming service has a dedicated service level agreements (SLAs) which can include the required reliability, end to end delay and the other QoS parameter. The second component is \emph{Infrastructure Network Provider (InP)} (InP) which is the owner of the commodity servers for running the VNFs and the links between the servers for routing the service's traffic. The last component is the \emph{Network Operator} (NO) which is responsible for responding to the incoming services according to their requested SLAs. For this purpose, NO should chain appropriate VNFs for each incoming service and then place them onto InP's servers [1].

One of the most important challenges in NFV is the placement of incoming services in the InP. In [7], the NFV placement problem with the purpose of energy and traffic cost minimization is considered. Also, they tried to prevent resource fragmentation in the servers. In [8], NFV placement problem is considered in a way that the cost of using servers and links is minimized and the requested delay of services is met. In [9], NFV placement problem with a cost function including deployment cost, resource cost, traffic cost, delay cost, and resource fragmentation cost is considered.  In [10]--[12], game theoretical models for NFV placement are considered. In [10],[11], the dynamic market mechanism design for on-demand service chain provisioning and pricing in the NFV market is studied. Authors in [12] model the selfish and competitive behavior of users in NFV with an atomic weighted congestion game is used.

In this paper, we consider a reliability-aware NFV placement problem. We would like to minimize the placement cost while maximizing the number of admitted services. A service will be in the perfect running state if all constituent VNFs of the service run in the commodity servers without failure. As a result, if only one of the servers which host the service VNFs fail, the service would be disrupted. We know that the servers in InPs can have different failure probabilities. In our work, NFV placement is carried out in an online manner. NO assigns the InP's servers to the incoming services according to the available resources. For the problem with dynamic characteristic, the learning-based approach like Q-learning can be useful [13]. Recently, Deep Q-Network (DQN) becomes useful because of some shortcomings in the Q-learning[14]. In this paper, we consider the use of DQN for NFV placement to meet the reliability requirement of the incoming services. In introduced NFV placement problem, NO learns the optimal policy in different states. The contributions of this paper are summarized as follows.

\begin{itemize}
\item We introduce an optimization problem for jointly minimizing the placement cost and maximizing the number of admitted services regarding the reliability requirement.
\item We introduce a solution based on DQN for reliability-aware NFV placement. For this purpose, we define the corresponding states and rewards of Q-Learning in an NFV framework.
\item Finally, we investigate the convergence of introduced DQN technique for NFV placement problem and evaluate the performance of this method concerning the admission ratio.
\end{itemize}

Machine Learning approaches have been used in NFV, recently. In [15], an efficient online algorithm from learning literature for dynamic placement of VNF service chains is presented. The considered objective function is operational cost minimization of the service chain provider. In [16], a machine-learning-based method for jointly optimization NFV placement and monitoring processes. In [17] by using Deep Feedforward Neural Network or Multi-Layer Perceptron (MLP), a solution for proactive identification of SLA violations is presented. Authors in [18] formulate the VNF selection and chaining problem as a Binary Integer Programming (BIP) model for end-to-end delay minimization. They propose a novel deep learning- based strategy for solving the problem.

The rest of the paper is organized as follows. We present the system model for reliable NFV placement problem in Section  \ref{Section: System model}. Then, we present a DQN model for reliable NFV placement problem in Section \ref{Section: DQN_NFV}. In Section \ref{Section: Numerical_Results}, we numerically evaluate the  performance  of  the  proposed  scenario. Finally, in  Section  \ref{Section: Conclusion}, we conclude the paper.

\section{System~Model}
\label{Section: System model}
We consider a scenario in which NO aims to deliver agile services using NFV. We assume there are multiple InPs with commodity servers that the NO can use them for placing the SFC of services. Each InP has some servers with a limited amount of resources. The main characteristic of each InP is the failure probability of its servers which is different from the other InPs. We assume that the unit server cost for each InP is dependent on the failure probability of InP. \\
\vspace{-6mm}
\subsection{Infrastructure Network Providers (InPs)}
Let $P$ denote the set of InPs and $S_i$ denote the set of servers of the $i^\text{th}$ InP. We indicate the number of the InPs with $|P|$ and the number of servers for the $i^{th}$ InP with $|S_i|$. We assume that the entire network of InPs can be shown with an undirected graph $G = (S, L)$ in which $S$ indicates the set of servers and $L$ indicates the set of the links between the servers as
\begin{align}
S &=\big\{S^m_i\mid m \in \{1,2,\ldots,|S_i|\}, i \in \{1,2,\ldots,|P|\}\big\} \label{Server_Set},
\end{align}
\vspace{-6mm}
\begin{align}
L =\big\{L^{mh}_{ij}\mid &m \in \{1,2,\ldots,|S_i|\}, \; h \in \{1,2,\ldots,|S_j|\},\notag\\
&i, j \in \{1,2,\ldots,|P|\}\big\},
\end{align}
where $S^m_i$ indicates the $m^\text{th}$ server of the $i^\text{th}$ InP and $L^{m,h}_{i,j}$ indicates the link between the $m^\text{th}$ server of the $i^\text{th}$ InP and the $h^\text{th}$ server of the $j^\text{th}$ InP. The resource amount of the $m^\text{th}$ server in the $i^\text{th}$ InP is denoted by $R_i^m$. The bandwidth of the link between the $m^\text{th}$ server in the $i^\text{th}$ InP and the $h^\text{th}$ server in the $j^\text{th}$ InP is denoted by $B_{i,j}^{m,h}$. The unit cost for using servers of the $i^\text{th}$ InP is denoted by $C_i$ and the unit cost for using the link between the $n^\text{th}$ server of the $i^\text{th}$ InP and the $m^\text{th}$ server of the $j^\text{th}$ InP is denoted by $C_{i,j}^{n,m}$. \\

Let $v_i$ indicate the failure probability of the servers of the $i^\text{th}$ InP. We assume that by decreasing the failure probability marginally close to zero, the unit cost for using the server is exponentially increased. Therefore, we consider an exponential model for the cost of using servers of different InP as
\begin{align}
C_i = \alpha e^{\beta(v_{\text{Base}}-v_i)}, \ i=1,\ldots,|P|,
\end{align}
\noindent where $\alpha$ and $\beta$ are design parameters and $v_{\text{Base}}$ is the highest acceptable failure probability. 

\subsection{Characteristics of Service Requests}

We divide the time into slots with equal length, and at the beginning of each slot, we consider the NFV placement problem for incoming service requests. Also, we assume that each service lasts for a random number of slots. The departure probability of an existing service in a slot is $d_l$, where $l$ indicates the type of service. According to this assumption, the departure probability of service in the $n^{\text{th}}$ slot is independent of the departure probability of this service in the $(n-1)^{\text{th}}$. As a result, the number of existing services in the network in the $n^{\text{th}}$ slot is only dependant on the number of existing services in the network in the $(n-1)^{\text{th}}$.

Let $L$ denotes the number of service types and $K_l$ indicates the number of requested services for the $l^{\text{th}}$ type of service in each slot. Also, the number of chained VNFs for the $l^{\text{th}}$ service type is $U_l$. We indicate the required bandwidth for this type of service with $b_l$ and the required resource of the $u^{\text{th}}$ VNF of this service type with $r^u_l$. It is worth noting that we consider only one resource type for a service. However the extension to the multiple resource type is straightforward. The maximum acceptable failure probability for the $l^{\text{th}}$ type of service is $F_l$. Finally, we indicate the decision variable of placing the $u^{\text{th}}$ VNF of the $(k_l)^{\text{th}}$ service of $l^{\text{th}}$ type in the $m^{\text{th}}$ server of the $i^{\text{th}}$ InP in the $n^{\text{th}}$ slot with $x_{i,k_l,n}^{m,u}\in\{0, 1\}$.

\subsection{Cost~Function}
The two main components of the cost function are server cost and link cost. Let $\xi^s_n$ denote the cost of using the servers in the $n^{\text{th}}$ slot. We can write $\xi^s_n$ as the summation of server cost for placement of each service type, $\xi^s_{n,k_l}$ as
\begin{align}
\xi^s_n = \sum_{l=1}^L\sum_{k_l=1}^{K_l}\xi^s_{n,k_l}, \; \xi^s_{n,k_l} = {\sum_{u=1}^{U_l}{\sum_{i=1}^{|P|}\sum_{m=1}^{|S_i|} x_{i,k_l,n}^{m,u} \times r_l^u\times C_i}}. 
\end{align}

It is worth noting that we assume the cost of using servers is a linear function of the binary decision variable, $x_{i,k_l,n}^{m,u}$. \\

The second main component of the cost function is the cost of using links between servers which is denoted by $\xi^l_n$. We can write the $\xi^l_n$ as the summation of server cost for placement of each service, $\xi^l_{n,k_l}$ as
\begin{align}
\xi^l_n &= \sum_{l=1}^L \sum_{k_l=1}^{K_l} \xi^l_{n,k_l}  \label{LinkCost} \\
\xi^l_{n,k_l} &= {\sum_{u=1}^{U_l-1}{\sum_{i=1}^{|P|}\sum_{m=1}^{|S_i|}\sum_{j=1}^{|P|}\sum_{h=1}^{|S_j|}
x_{i,k_l,n}^{m,u}\times x_{j,k_l,n}^{h,u+1} \times b_l \times C_{i,j}^{m,h}}} \notag,
\end{align}
where $x_{i,k_l,n}^{m,u}\times x_{j,k_l,n}^{h,u+1}$ is used to indicate the use of the link between the $m^{\text{th}}$ server of the $i^{\text{th}}$ InP and the $h^{\text{th}}$ server of the $j^{\text{th}}$ InP, for forwarding of the traffic between $u^{\text{th}}$ and $(u+1)^{\text{th}}$ VNFs of $k_l^{\text{th}}$ service of $l^{\text{th}}$ type. It is worth noting that if two consecutive VNFs of a service are placed in the same server ($m=h$ and $i=j$), then $C_{i,j}^{m,h} = 0$ and there is no cost for forwarding of traffic between these VNFs. As seen in (\ref{LinkCost}), this cost component is a nonlinear function of the binary decision variable, $x_{i,k_l,n}^{m,u}$. The total cost in the $n^{\text{th}}$ slot is
$\xi^T_n=\xi^s_n+\xi^l_n$. 

\subsection{Reliability~Constraint}

We indicate the failure probability for the $k_l^{\text{th}}$ service of the $l^{\text{th}}$ type in the $n^{\text{th}}$ slot with $f_{k_l,n}$. The reliability constraint is $f_{k_l,n} \leq F_l$. To obtain $f_{k_l,n}$, we should calculate the probability of being in the running state (i.e., not being failed) for this service, $p_{k_l,n}$. We know that a service is in running state if all VNFs of service are not failed. As a result, we should determine the failure probability of a VNF which is a function of the binary decision variable, $x_{i,k_l,n}^{m,u}$. Let $f^{u}_{k_l,n}$ denote the failure probability of the $u^{\text{th}}$ VNF for the $(k_l)^{\text{th}}$ service of the $l^{\text{th}}$ type in the $n^{\text{th}}$ slot. This probability is calculated as
\vspace{-2mm}
\begin{align}
f_{k_l,n}^{u} = \prod_{i=1}^{|P|}\big(\prod_{m=1}^{|S_i|} \rho_{i,k_l,n}^{m,u}\big), \; \rho_{i,k_l,n}^{m,u} = \begin{cases} v_i,   & x_{i,k_l,n}^{m,u} = 1 \\ 1,   & x_{i,k_l,n}^{m,u} = 0 \end{cases}, \label{eq:VNF_Error_Probability}
\end{align}
where $v_i$ is the failure probability of the $i^{\text{th}}$ InP. According to (\ref{eq:VNF_Error_Probability}), the failure probability for the $u^{\text{th}}$ VNF of the $(k_l)^{\text{th}}$ service of the $l^{\text{th}}$ type in the $n^{\text{th}}$ slot is the multiplication of the failure probability of the VNF in all InPs. Also, the failure probability of a VNF in each InP is the multiplication of failure probability of all the servers in which the respective VNF is placed. We assume failure events in different InPs and also in different servers of an InP are independent.
We calculate the probability of being in the running state for the $(k_l)^{\text{th}}$ service of $l^{\text{th}}$ type in the $n^{\text{th}}$ slot as $p_{k_l,n} = \prod_{u=1}^{U_l}{(1-f_{k_l,n}^{u})}$. Finally, we can calculate the failure probability for the $(k_l)^{\text{th}}$ service of $l^{\text{th}}$ type in the $n^{\text{th}}$ slot as $f_{k_l.n} = 1-p_{k_l,n}$.

\subsection{Minimum Cost NFV Placement}

In this part, we want to formulate the objective of NO throughout the time as an optimization problem. We assume that the purpose of NO is minimizing the placement cost regarding the reliability requirement of incoming services and InP's resource constraint. Thus, the optimization problem can be written as
\vspace{-2mm}
\begin{align}
\min_{x_{i,k_l,n}^{m,u}} &\quad \textstyle \sum_{n=0}^{\infty} \gamma^n \xi^n_T \label{eq:Objective_11} \quad \text{s. t.}\\
&\hspace{-1cm}\textstyle \sum_{i=1}^{|P|}{\sum_{m=1}^{|S_i|}{x_{i,k_l,n}^{m,u}} = 1} \label{eq:Constraint_11}\\
&\hspace{-1cm}\textstyle \prod_{u=1}^{U_l}{\bigg(1-\prod_{i=1}^{|P|}\Big(\prod_{m=1}^{|S_i|} \rho_{i,k_l,n}^{m,u} \Big)\bigg)} \geq \big(1-F_l\big) \label{eq:ReliabilityCons}\\
&\hspace{-1cm}\textstyle \sum_{l=1}^{L}\sum_{k_l=1}^{K_l}{\sum_{u=1}^{U_l}{x_{i,k_l,n}^{m,u} \times r_l^u \leq R_{i,n}^m}} \label{eq:Constraint_12}\\
&\hspace{-1cm}\textstyle \sum_{l=1}^{L}\sum_{k_l=1}^{K_l}{\sum_{u=1}^{U_l-1}{x_{i,k_l,n}^{m,u}\times x_{j,k_l,n}^{h,u+1} \times b_l \leq B_{i,j,n}^{m,h}}} \label{eq:Constraint_13} \\
&\hspace{-1cm}x_{i,k_l,n}^{m,u} \in \{0, 1\}, i,j =1,\ldots,|P|, \,  m =1,\ldots,|S_i|, \label{eq:Constraint_14}\\ \notag
&\hspace{-1cm}\, h =1,\ldots,|S_j|, u =1,\ldots,U_l, \, k_l =1,\ldots, K_l, \\ \notag
&\hspace{-1cm}\, l = 1,\ldots, L, n=0,1,2,3,\ldots \notag
\end{align}
where $R_{i,n}^m$ indicates the remaining resource for the $m^{\text{th}}$ server of the $i^{\text{th}}$ InP in the $n^{\text{th}}$ slot and $B_{i,j,n}^{m,h}$ is the remaining bandwidth for the link between the $m^{\text{th}}$ server of the $i^{\text{th}}$ InP and the $h^{\text{th}}$ server of the $j^{\text{th}}$ InP, in the $n^{\text{th}}$ slot. Constraint in (\ref{eq:Constraint_11}) indicates that each VNF is instantiated once. Constraint in (\ref{eq:ReliabilityCons}) guarantees the reliability requirement of each service. The constraint in (\ref{eq:Constraint_12}) makes sure that the resource capacity of each server is not violated in each slot. The constraint in (\ref{eq:Constraint_13}) guarantees that the bandwidth capacity of each link is not violated in each slot.

The optimization problem in  (\ref{eq:Constraint_11})-(\ref{eq:Constraint_14}) is intractable for large networks with various services. Learning based techniques can be helpful to solve such problem. The goal of the learning technique is to learn a policy which determines what action to take in each state. In the following, we introduce a model based on DQN for NFV placement problem regarding the reliability requirement.

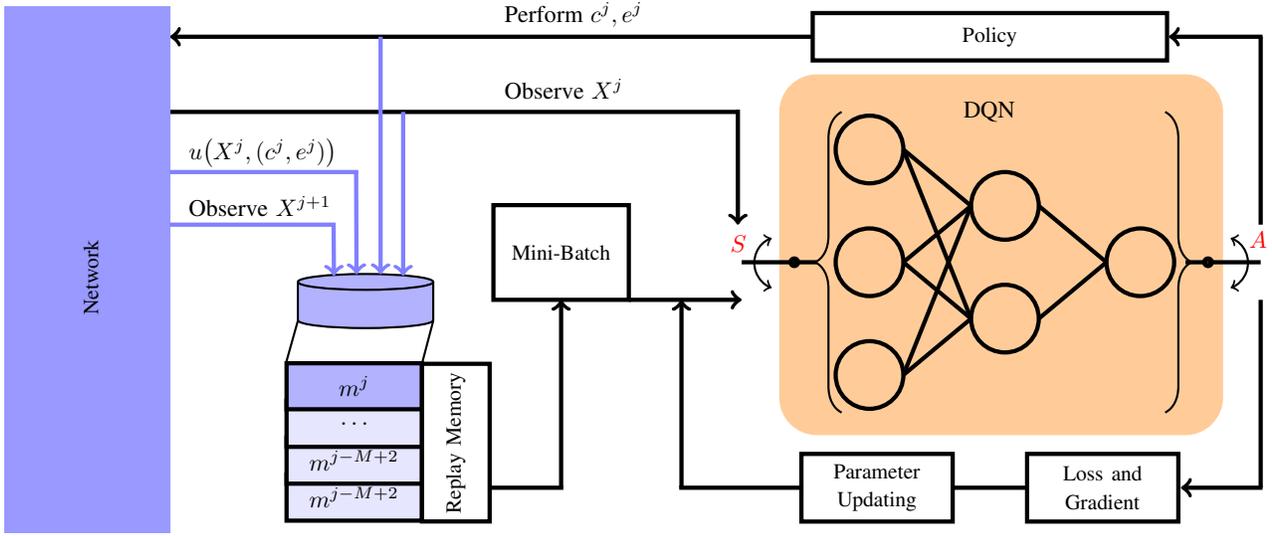
\begin{figure*}[!t]
\centering
\begin{tikzpicture}[scale=1, every node/.style={scale=0.9}]
\draw (8.84,4) circle (0.001cm);
\fill[orange!40!white, rounded corners=5mm] (19.3,5.7) rectangle (25.2,10.5);
\node[text width=3cm] at (23.1,10) {{DQN}};

\draw [thick,decorate,decoration={brace,amplitude=10pt},xshift=-2pt,yshift=0pt] (20.2,6) -- (20.2,10.0);
\draw [ultra thick] (19.5,8) -- (19.8,8);
\draw [ultra thick] (19.5,8) circle (0.05cm);
\draw [ultra thick] (19.5,8) -- (18.8,8);

\draw [thick,decorate,decoration={brace, mirror, amplitude=10pt},xshift=-2pt,yshift=0pt] (24.5,6) -- (24.5,10.0);
\draw [ultra thick] (24.7,8) -- (25,8);
\draw [ultra thick] (25,8) circle (0.05cm);
\draw [ultra thick] (25,8) -- (25.7,8);

\draw [ultra thick] (20.5,6.5) circle (0.45cm);
\draw [ultra thick] (20.95,6.5) -- (21.85,7.25);
\draw [ultra thick] (20.95,6.5) -- (21.85,8.75);

\draw [ultra thick] (20.5,8) circle (0.45cm);
\draw [ultra thick] (20.95,8) -- (21.85,7.25);
\draw [ultra thick] (20.95,8) -- (21.85,8.75);

\draw [ultra thick] (20.5,9.5) circle (0.45cm);
\draw [ultra thick] (20.95,9.5) -- (21.85,7.25);
\draw [ultra thick] (20.95,9.5) -- (21.85,8.75);

\draw [ultra thick] (22.3,7.25) circle (0.45cm);
\draw [ultra thick] (22.75,7.25) -- (23.65,8);

\draw [ultra thick] (22.3,8.75) circle (0.45cm);
\draw [ultra thick] (22.75,8.75) -- (23.65,8);

\draw [ultra thick] (24.1,8) circle (0.45cm);

\draw [ultra thick] (25.7,8.5) -- (25.7,11);
\draw [<-,ultra thick] (24.48,11) -- (25.725,11);
\draw [ultra thick] (25.7,7.5) -- (25.7,5);
\draw [<-,ultra thick] (24.64,5) -- (25.725,5);
\node[draw, ultra thick, align = center,text width=2cm, minimum height=1cm] at (23.6,5) {\small{Loss and Gradient}};
\draw [ultra thick] (22.6,5) -- (21.6,5);
\node[draw, ultra thick, align = center,text width=2cm, minimum height=1cm] at (20.6,5) {\small{Parameter Updating}};
\draw [<-,ultra thick] (18,7.5) -- (18,5);
\draw [ultra thick] (17.97,5) -- (19.62,5);
\node[draw, ultra thick, align = center,text width=5cm,minimum height=0.7cm] at (22.1,11) {\small{Policy}};
\draw [->,ultra thick] (19.75,11) -- (11.2,11);
\node[text width=3cm] at (17,11.3) {Perform $c^j,e^j$};
\draw [ultra thick] (18.75,10) -- (11.2,10);
\draw [->,ultra thick] (18.75,10.03) -- (18.75,8.5);
\node[text width=3cm] at (17,10.3) {Observe $X^j$};

\node[draw, ultra thick, fill=blue!10, align = center, minimum width=2cm, minimum height=0.68cm] at (13.65,4.868) {$m^{j-M+2}$};
\node[draw, ultra thick, fill=blue!10, align = center, minimum width=2cm, minimum height=0.68cm] at (13.65,5.36) {$m^{j-M+2}$};
\node[draw, ultra thick, fill=blue!10, align = center, minimum width=2cm, minimum height=0.68cm] at (13.65,5.85) {$\ldots$};
\node[draw, ultra thick, fill=blue!30, align = center, minimum width=2cm, minimum height=0.68cm] at (13.65,6.35) {$m^j$};
\node[draw, ultra thick, align = center, rotate=90, minimum width=2cm, minimum height=1cm] at (15,5.6) {\small{Replay Memory}};
\draw [<-,ultra thick] (16.4,7.5) -- (16.4,5);
\draw [ultra thick] (15.44,5) -- (16.43,5);
\node[draw, ultra thick, align = center,minimum width=2cm, minimum height=1.4cm] at (16.4,8.13) {\small{Mini-Batch}};
\draw [->,ultra thick] (17.3,7.5) -- (18.8,7.5);

\node [cylinder, fill=blue!30, thick, shape border rotate=90, draw,minimum height=0.8cm,minimum width=2cm] at (13.8,7.4){};
\draw [[->,ultra thick, blue!50] (14,11) -- (14,7.84);
\draw [[->,ultra thick, blue!50] (14.3,10) -- (14.3,7.82);
\draw [thick] (12.74,6.65) -- (12.9,7.22);
\draw [thick] (14.55,6.65) -- (14.7,7.22);

\fill[blue!40!white] (9,4.4) rectangle (11.2,11.4);
\node[rotate=90, minimum width=2.4cm, minimum height=1cm] at (10.15,7.8) {\small{Network}};

\draw [ultra thick, blue!50] (11.2,9.2) -- (13.7,9.2);
\draw [->,ultra thick, blue!50] (13.675,9.22) -- (13.68,7.84);
\node[text width=3cm] at (12.8,9.45) {$u\big(X^j,(c^j,e^j)\big)$};

\draw [ultra thick, blue!50] (11.2,8.5) -- (13.4,8.5);
\draw [[->,ultra thick, blue!50] (13.375,8.52) -- (13.38,7.82);
\node[text width=3cm] at (12.8,8.75) {Observe $X^{j+1}$};

\draw [<->, thick] (25.3,8.32) arc (70:-70:10pt);
\node[text width=1cm, red] at (26, 8.3) {{$A$}};
\draw [<->, thick] (19.2,8.32) arc (110:250:10pt);
\node[text width=1cm, red] at (19.1, 8.25) {{$S$}};

\end{tikzpicture}
\label{DQN_Figure1}
\caption{An overview of a Deep Q-Learning Neural Network (DQN).}
\end{figure*}

\section{DQN Model for NFV Placement}
\label{Section: DQN_NFV}
In this section, we introduce a model based on DQN for NFV placement considering the reliability requirement of the incoming services. First of all, we review Q-Learning and motivation for the combination of Q-Learning and Deep Neural Network (DNN). Then, we introduce a DQN model for NFV placement regarding the reliability.
\subsection{DQN Background}
In Reinforcement Learning, there are some agents who explore and exploit the environment based on the reward gained from an environment and the state which encapsulate all features and conditions by using a particular policy. The policy is used to make a balance between the exploration and the exploitation of agents. Rewards are the direct consequence of actions made by the agent in each state. Despite all the merits provided by Q-learning, its weak point lies in decision making in the problems where states are covering a wide range of possibilities and Q-tables are large. DQN is a combination of both neural networks and Q-learning approaches. DQN uses the same model but instead of updating the Q-table which is hard to be searched in environments with big states space, it trains a DNN while it explores and exploits. By making each action, the reward gained by the agent is used to conduct the back-propagation process and update the weights of neural networks. The input of the neural network is a vector representing the state, and the possible actions are the output neurons of the neural network which are selected by the agent based on a policy.

The general overview of the DQN-agent used in this approach is shown in Fig. 1. The states are given as inputs of the DNN, and all possible actions are at the output of the neural network. The chosen actions based on the policy affects the NFV network. The environment returns the direct consequence as specific rewards to the memory. The states, next states, chosen actions, and rewards of all slots are stored in the memory. A mini-batch is randomly sampled from memory for updating the weights of the neural network.\\
\vspace{-5mm}
\subsection{Modeling NFV Placement with DQN}

For DQN problem, we characterize a four-tuple including state set, action set, reward set and memory set. We show these four-tuple with
$(\Omega_S, \Omega_A, \Omega_R, \Omega_M)$ where $\Omega_S$ is the state set, $\Omega_A$ is the action set, $\Omega_R$ is the reward set and $\Omega_M$ is the memory segment. For NFV placement, we take a new approach towards defining the states. In large-scale problems, choosing states in a way that represents our demands to the network is crucial. Our most prior goal is to satisfy the requested reliability of each service in each slot. Thus, the trained system should discriminate between services with variate reliability requirements. Our DQN agent should also be aware of the available resources that each InP can provide at the moment of decision for each incoming service placement. It is worth noting that discrimination among services should be considered by the DQN agent, as resources demanded among two services may differ and its the networks duty to choose the best corresponding VNFs to satisfy these resource demands.

The decision for selecting the InPs to allocate resources to an incoming service should be made by considering the amount of resource needed while taking available resources distributed among InPs and the required reliability in mind. Combination of all these prerequisites generates a complex and large space state that our DQN agent should be able to comprehend through learning and iteration. \\

We define states as a vector of available resources provided by InPs concatenated with the resources demanded by a service, and the requested reliability that should be satisfied. As a result, the state set of DQN agent can be written as
\begin{align}
&\Omega_S = \Big\{\omega_i^m, r_l^u, F_l\Big\}, \; \; \textstyle l \in {1, 2, \ldots, L}, \; \; 1 \leq u \leq U_l, \\
&\textstyle 1 \leq m \leq |S_i|, \; \; 1 \leq i \leq |P|, \; \; 0 \leq \omega_i^m \leq R_i^m \notag 
\end{align}
\noindent where  $\omega_i^m$ indicates the remaining resources for the $i^{\text{th}}$ server of the $m^{\text{th}}$ InP, $r_l^u$ denotes the demand resource for the $u^{\text{th}}$ VNF of the $l^{\text{th}}$ service type and $F_l$ is the reliability requirement of the $l^{\text{th}}$ service type. It is worth noting that the resource demand for each service is considered in the state to highlight the characteristics of the incoming services for the learning agent.\\
\vspace{-0.4mm}
Let $K$ indicate the maximum possible number of service requests in each slot. $K$ is chosen according to the possible resource budget of the InPs. In DQN modeling of NFV placement, we assume that in each slot, the NO considers the placement of service requests one by one. More precisely, the placement of the first incoming service is determined, the DQN states are updated, then the placement of the second incoming service is determined and so on. We define the action as the possible placement policies for each VNF of an incoming service which can be written as
$\Omega_A = \big\{(a,b) \big|\; 1 \leq a \leq |P|, \; 1 \leq b \leq |S_a| \big\},$
\noindent where $a$ indicates the InP index and $b$ denotes the server index in the respective InP. The learning agent uses the DQN outputs to determine the InP and server index for all VNFs of a service. We consider $N$ outputs for the DQN as $Q_{\text{Out}} = \big\{q_1, q_2, \ldots, q_N\big\}$ in which $q_i$ indicates the q-value for assigning the corresponding server of the $i^{\text{th}}$ output to a VNF of the considered service. The value of $N$ depends on the number of InP, number of servers of each InP, resource budget of each server and the maximum value for the resource demands of a VNF for all service types. For a service with $U_l$ VNFs, NO determines the placement using $Q_{\text{Out}}$. The optimal solution for a service with $U_l$ VNFs is to select $U_l$ servers with the highest Q-values. However, at the beginning, the agent has no sense about the optimal solution. As a result, the learning agent needs to consider all possible actions. In the RL, $\epsilon$ gives the agent a chance of exploration.

Due to the quiddity of the Q-Learning and particularly DQN, defining the best reward and cost functions play a vital role. The most important task for a NO in the placement of the services is to meet the requested reliability. We outline a penalty for a situation in which the reliability requirement is not satisfied. On the other hand, we assign a reward for the successful placement of a service. However, due to the limited resource budget of InPs, the agent should comprehend to provide reliability as close as possible to the requested reliability. This proximity should be implemented in the structure of reliability rewards. Because of the nature of the learning system, it is possible that the selected server for hosting a VNF does not have enough resource. We outline a penalty for situations in which resource allocation is failed due to the lack of enough resource in the selected server. Finally, we consider the placement cost of the allocated resources for an incoming service as a penalty term. Now, the placement reward $R_p$, for a service with type $l$, is written as
\begin{align}
R_p = \begin{cases}
-V_R \, & \exists u \leq U_l, \, R_{a_u}^{b_u} \leq \omega_{a_u}^{b_u} + r_l^u  \\
-V_F \, & \nexists u \leq U_l, \, R_{a_u}^{b_u} \leq \omega_{a_u}^{b_u} + r_l^u \; \text{and} \; F_l \leq F_p \\
V_S - V_p \, &\nexists u \leq U_l, \, R_{a_u}^{b_u} \leq \omega_{a_u}^{b_u} + r_l^u \; \text{and} \; F_l \geq F_p
\end{cases}
\end{align}
\vspace{-8mm}
\begin{align}
V_p = &\sum_{u=1}^{U_l-1}{b_l \times C_{a_u,a_{u+1}}^{b_u,b_{u+1}}} + \sum_{u=1}^{U_l}{r_l^u \times C_{a_u}}\\
V_S = &R_S \times \exp\big\{-(F_l-F_p)\big\}, \label{ReliabilityReward} \\
A_p = &\Big\{ (a_u,b_u) \big| 1 \leq u \leq U_l \Big\},\\
F_p = &1 - \prod_{u=1}^{U_l}{(1-v_{a_u})},
\end{align}
where $A_p$ indicates the output action of the DQN in which $a_u$ and $b_u$ denote the allocated InPs and server index for the $u^{\text{th}}$ VNF of the service, $F_l$ is the maximum acceptable failure probability of the $l^{\text{th}}$ service type, $F_p$ denotes the failure probability of placement. The value of the $V_F$ is a penalty for not admitting the service because of the reliability, and $V_R$ is a penalty for violating the resource budget of the selected servers. $V_p$ is the placement cost including the server and link costs. Finally, $V_S$ is a reward for the successful placement of the service. According to (\ref{ReliabilityReward}), we consider more reward for the placement in which the placement reliability is in the proximity of requested reliability.

We defined a structure as a memory for the agent so that through the learning, the agent would not tend to adapt to a sequence of services and be able to work on a random batch of the services and try to improve the previous acquired results based on the new knowledge gained through experience [14]. For this purpose, we use the current state, action, reward and next state as a memory segment of the DQN agent for each service after its allocation. When the number of considered services reaches a threshold, a random batch of memory segments is selected for updating weights of the neural network.
\vspace{-2mm}
\section{Numerical Results}
\label{Section: Numerical_Results}
In this section, we evaluate the performance of the proposed scheme regarding the total placement cost and the admission ratio. For the simulation framework of the DQN, we used Keras and TensorFlow in Python. For the InPs, we consider seven InPs with different radiabilities $[96, 97, 98, 99, 99.9]$. We assume each InP has five servers with the same reliability level. For each server, we consider the capacity of 100 units of one resource type. We consider five service types. The service type requested reliability is assumed among $[91, 92, 93, 94, 95]$, according to the SLA requirement of Google Apps [19]. Also, we assume that the number of VNFs in each service type is between three to five VNFs and the resource demand of the VNFs is considered to be between 10 and 20 units. We assume that the departure probability for all service types is equal and between $0.6$ to $0.8$. For the DQN network, we use a fully connected DNN, which involves hyperbolic tangent and ReLu as activation function in the middle layer, and the output layer is connected to a linear activation function [20]. Each layer is associated with Dropout, with its parameter set between 0.05 and 0.2, so that overfitting is prohibited [21]. Also, we use the mean square error (MSE) metric for error function.

\begin{figure}[!t]
\center
\includegraphics[scale=0.38]{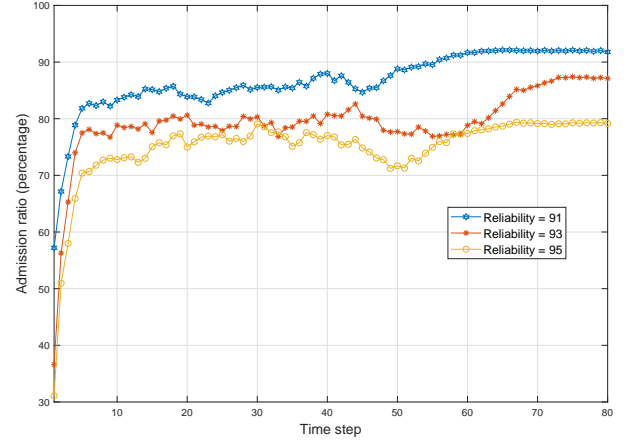}
\centering
\caption{\footnotesize{Admission ratio for different service types during the learning.}}
\label{Admission_ratio_Learning}
\end{figure}
\begin{figure}[!t]
\centering
\center
\includegraphics[scale=0.38]{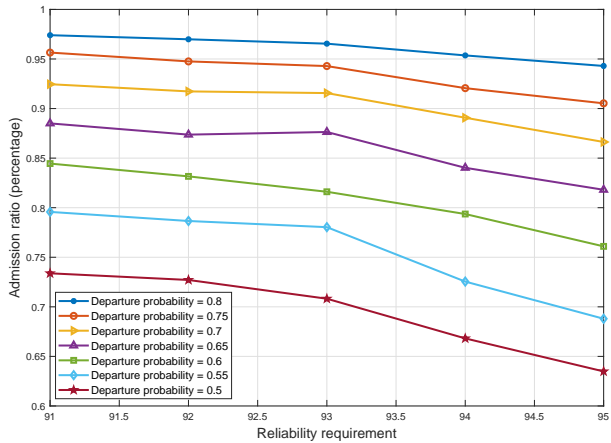}
\caption{\footnotesize{Admission ratio for different service types and departure probabilities.}}
\label{Admission_ratio_Departure_Prob}
\end{figure}


For evaluating the performance of the proposed DQN-agent, we consider the admission ratio for different service types. The admission ratio for each service type is defined as the proportion of the successfully accepted services regarding the reliability requirements to the number of incoming services. The trend of learning and adaptation of the DQN-agent for admitting services with different reliability requirements through time steps is shown in Fig. \ref{Admission_ratio_Learning}. The x-axis shows the time steps which each consists of 10000 slots. The y-axis indicates the admission ratio for different reliability requirement. The agent policy is $\epsilon-$greedy which initiates by the value of 0.2 to strengthen the aspect of exploration in initial states of learning. As time goes by, our agent tends to be more exploitative rather than being explorative due to our decaying $\epsilon$, which results in consistent values. For high-reliability requirements, the convergence time of the agent increases and the value which each service type converges to is decreased.


For evaluating the robustness of the proposed DQN method for dynamic reliability-aware NFV placement, we consider the performance of the DQN-agent under different departure probabilities. For this purpose, we use the optimal policy for placement of the incoming services. The resulted admission ratio for different values of the departure probabilities for a fixed resource amount of the InPs is shown in Fig. \ref{Admission_ratio_Departure_Prob}. As seen in Fig. \ref{Admission_ratio_Departure_Prob}, with a decrease in the value of the departure probability, the admission ratio is decreased.

\section{Conclusion}
\label{Section: Conclusion}
In this paper, we considered a dynamic reliability-aware NFV placement for NFV-enabled NO using DQN. For this purpose, we considered a multi-InP scenario in which different levels of reliability with different costs are offered to the NO. On the other hand, we considered multiple service types for the incoming services which introduced by their reliability requirement. Also, we assumed that admitted services would be ended in each slot with a departure probability which can be different for various service types. For DQN-agent, we defined the state set, action set, reward and memory considering the objective of the NO which is maximizing the admission ratio while minimizing the placement cost. Using simulations, we showed that the NO could learn how to effectively use the resources of the InPs for various service types in different states in a way that the admission ratio is maximized and placement cost is minimized.

\section{References}

\begin{enumerate}
	
	\scriptsize{\item[{[1]}] J. G. Herrera and J. F. Botero, ``Resource allocation in {NFV}: A comprehensive
		survey,'' \emph{IEEE Trans. on Network and Service Management}, vol. 13,
		no. 3, pp. 518--532, 2016.}

	\item[{[2]}] S. Khebbache, M. Hadji, and D. Zeghlache, ``Virtualized network functions
	chaining and routing algorithms,'' \emph{Computer Networks}, vol. 114, pp.
	95--110, 2017.
	
	\item[{[3]}] J. Sherry, S. Ratnasamy, and J. S. At, ``A survey of enterprise middlebox
	deployments,'' Technical Report UCB/EECS-2012-24, EECS Department, University
	of California, Berkeley, 2012.
	
	\item[{[4]}] X. Zhang, C. Wu, Z. Li, and F. C. Lau, ``Proactive {VNF} provisioning with
	multi-timescale cloud resources: Fusing online learning and online
	optimization,'' in \emph{IEEE INFOCOM}, Atlanta, GA, May. 2017.
	
	\item[{[5]}] G.~NFV, ``Network functions virtualisation {NFV}; architectural framework,''
	\emph{NFV ISG}, Oct. 2013.

	\item[{[6]}] R. Mijumbi, J. Serrat, J.-L. Gorricho, N. Bouten, F. De Turck, and R. Boutaba,
	``Network function virtualization: State-of-the-art and research
	challenges,'' \emph{IEEE Communications Surveys \& Tutorials}, vol. 18,
	no. 1, pp. 236--262, 2016.
	
	\item[{[7]}] C.~Pham, N.~H. Tran, S.~Ren, W.~Saad, and C.~S. Hong, ``Traffic-aware and
	energy-efficient {VNF} placement for service chaining: Joint sampling and
	matching approach,'' \emph{IEEE Trans. on Services Computing}, 2017.
	
	\item[{[8]}] M.~Mechtri, C.~Ghribi, and D.~Zeghlache, ``A scalable algorithm for the
	placement of service function chains,'' \emph{IEEE Trans. on Network and
		Service Management}, vol.~13, no.~3, pp. 533--546, 2016.
	
	\item[{[9]}] F.~Bari, S.~R. Chowdhury, R.~Ahmed, R.~Boutaba, and O.~C. M.~B. Duarte,
	``Orchestrating virtualized network functions,'' \emph{IEEE Trans. on Network
		and Service Management}, vol.~13, no.~4, pp. 725--739, 2016.
	
	\item[{[10]}] S.~Gu, Z.~Li, C.~Wu, and C.~Huang, ``An efficient auction mechanism for service
	chains in the {NFV} market,'' in \emph{IEEE INFOCOM}, San Fransisco, CA, May.
	2016.
	
	\item[{[11]}] X.~Zhang, Z.~Huang, C.~Wu, Z.~Li, and F.~C. Lau, ``Online stochastic buy-sell
	mechanism for {VNF} chains in the {NFV} market,'' \emph{IEEE Journal on
		Selected Areas in Communications}, vol.~35, no.~2, pp. 392--406, 2017.
	
	\item[{[12]}] S.~D’Oro, L.~Galluccio, S.~Palazzo, and G.~Schembra, ``Exploiting congestion
	games to achieve distributed service chaining in {NFV} networks,'' \emph{IEEE
		Journal on Selected Areas in Communications}, vol.~35, no.~2, pp. 407--420,
	2017.

	\item[{[13]}] C.~J. C.~H. Watkins, ``Learning from delayed rewards,'' Ph.D. dissertation,
	King's College, Cambridge, UK, May, 1989.
	
		\item[{[14]}] V.~Mnih, K.~Kavukcuoglu, D.~Silver, A.~Graves, I.~Antonoglou, D.~Wierstra, and
		M.~Riedmiller, ``Playing atari with deep reinforcement learning,''
		\emph{arXiv preprint arXiv:1312.5602}, 2013.

		\item[{[15]}] Y.~Jia, C.~Wu, Z.~Li, F.~Le, A.~Liu, Z.~Li, Y.~Jia, C.~Wu, F.~Le, and A.~Liu,
		``Online scaling of {NFV} service chains across geo-distributed
		datacenters,'' \emph{IEEE/ACM Trans. on Networking (TON)}, vol.~26, no.~2,
		pp. 699--710, 2018.

		\item[{[16]}] V.~Sciancalepore, F.~Z. Yousaf, and X.~Costa-Perez, ``{z-TORCH}: An automated
		{NFV} orchestration and monitoring solution,'' \emph{IEEE Trans. on Network
			and Service Management}, 2018.

		\item[{[17]}] J.~Bendriss, I.~G.~B. Yahia, R.~Riggio, and D.~Zeghlache, ``A deep learning
		based sla management for {NFV}-based services,'' in \emph{Conference on
			ICIN}, Paris, France, Feb. 2018.
	
			\item[{[18]}] J.~Pei, P.~Hong, and D.~Li, ``Virtual network function selection and chaining
			based on deep learning in sdn and {NFV}-enabled networks,'' in \emph{ICC
				Workshops}, Kansas City, MO, USA, May. 2018.
	
			\item[{[19]}] ``Google apps service level agreement,'' [Online].
			Available:http://www.google.com/apps/intl/en/terms/sla.html.

			\item[{[20]}] A.~L. Maas, A.~Y. Hannun, and A.~Y. Ng, ``Rectifier nonlinearities improve
			neural network acoustic models,'' in \emph{ICML}, vol.~30, no.~1, 2013, p.~3.

			\item[{[21]}] N.~Srivastava, G.~Hinton, A.~Krizhevsky, I.~Sutskever, and R.~Salakhutdinov,
			``Dropout: a simple way to prevent neural networks from overfitting,''
			\emph{The Journal of Machine Learning Research}, vol.~15, no.~1, pp.
			1929--1958, 2014.

\end{enumerate}

\end{document}